\documentclass[%
reprint,
superscriptaddress,
 amsmath,amssymb,
 aps,
showkeys
]{revtex4-2}

\usepackage[usenames, dvipsnames]{color}
\usepackage{xcolor}
\usepackage{graphicx}
\usepackage{dcolumn}
\usepackage{bm}

\usepackage{hyperref}

\begin{document}

\title{Partial Charge Transfer and Absence of Induced Magnetization in EuS(111)/Bi$_2$Se$_3$ Heterostructures}

\author{Damien Tristant}
\affiliation{Cain Department of Chemical Engineering, Louisiana State University, Baton Rouge, Louisiana 70803, United States of America.}

\author{Ilya Vekhter}
\email{vekhter@lsu.edu}
\affiliation{Department of Physics and Astronomy, Louisiana State University, Baton Rouge, Louisiana 70803, United States of America.}

\author{Vincent Meunier}
\affiliation{Department of Physics, Applied Physics, and Astronomy, Rensselaer Polytechnic Institute, Troy, New York 12180, United States of America.}

\author{William Shelton}
\email{wshelton@lsu.edu}
\affiliation{Cain Department of Chemical Engineering, Louisiana State University, Baton Rouge, Louisiana 70803, United States of America.}

\date{\today}

\begin{abstract}

Heterostructures made from topological and magnetic insulators promise to form excellent platforms for new electronic and spintronic functionalities mediated by interfacial effects. We report the results of a first-principles density functional theory study of the geometric, electronic structure, and magnetic properties of EuS(111)/Bi$_2$Se$_3$ interface, including van der Waals and relativistic spin-orbit effects. In contrast to previous theoretical studies, we find no appreciable magnetic anisotropy in such a heterostructure. We also do not see additional induced magnetization at the interface or the magnetic proximity effect on the topological states. This is due to the localized nature of Eu moments, and because of a partial charge transfer of $\sim$0.5 electron from Eu to Se. The formation of the surface dipole shifts the Dirac cone about 0.4~eV below the chemical potential, and the associated electrostatic screening moves the topological state from the first to the second quintuple layer of Bi$_2$Se$_3$.  

\end{abstract}


\maketitle

\section{INTRODUCTION}

Topological insulators (TIs) are a class of materials with unique quantum-mechanical properties that could have a significant impact on next-generation spintronic and quantum information devices.~\cite{essin2009magnetoelectric,yu2010quantized,li2010dynamical,ferreira2013magnetically,fan2014magnetization,fan2016electric}  They are insulating in the bulk, while their surfaces support metallic two-dimensional states.~\cite{hasan2010colloquium,qi2011topological} These surface states, characterized by a massless Dirac dispersion, are topologically protected by time-reversal symmetry. Their spin is locked in the direction perpendicular to the momentum, and, for a high-symmetry surface orientation, are in the plane of the interface. 

When the time-reversal symmetry is broken following the application of an external or exchange magnetic field normal to the surface, the surface states become massive,~\cite{chen2010massive} and the resulting gapped topological band structure supports, \textit{e.g.}, a quantum anomalous Hall (QAH) effect.~\cite{Wang_2015,yu2010quantized,chang2013experimental} 

The most studied TI materials are members of the sesqui-chalcogenides family, \textit{e.g.}, Bi$_2$Te$_3$, Sb$_2$Te$_3$, and Bi$_2$Se$_3$, where bismuth selenide has the largest measured bulk band gap of $\sim$0.30 eV.~\cite{hermanowicz2016topological} Imposing an exchange field on the surface states has been pursued either by doping or substituting cations in the bulk TIs with magnetic dopants, such as Fe, Mn, and Cr atoms.~\cite{figueroa2015local,collins2014magnetic} In addition, to preserve the structural integrity of TIs, the concentration of dopants, \textit{e.g.}, Mn atoms, should not exceed 7.5$\%$.~\cite{figueroa2015local} However, controlling the diffusion of these elements, as well as the magnetic order, remains a challenge. 

The alternative approach to impose an exchange field is to interface TIs with a structurally matching  magnetically ordered material. Europium sulfide (EuS) emerged as a strong candidate for such interfaces due to its large ordered moment and band gap, as well as a good match along the (111) direction of its rock salt lattice to the lattice constant of Bi$_2$Se$_3$. However, so far both available experimental results and published theoretical predictions for such interfaces have failed to establish a consensus on the viability of these heterostructures for the realization of QAH and related phenomena. 

The key issues are whether the Eu spins tilt from the bulk direction parallel to the interface to enable the opening of the gap in the surface TI state, whether there is a strong enough hybridization between EuS and Bi$_2$Se$_3$ wave function to ensure magnetic proximity effect, and whether the topological states remain near the chemical potential once the interface is formed. There is evidence for the in-plane components of the magnetic polarization,~\cite{wei2013exchange,lee2016direct,katmis2016high} at least in thin film geometries. However, while early results suggested significant proximity coupling to the surface states,~\cite{katmis2016high} subsequent measurements indicated that the magnetic field penetration can be either comparable to that in a non-topological material~\cite{krieger2019topology} or absent altogether.~\cite{figueroa2020absence,meyerheim2020structure} Raman measurements~\cite{osterhoudt2018charge} showed charge transfer at the interface, consistent with \textit{ab initio} density functional theory (DFT) based calculations for a fixed out-of-plane orientation of Eu moments.~\cite{lee2014magnetic} The magnetic anisotropy, however, was found to be strongly strain-dependent in another theoretical study,~\cite{kim2017understanding} which predicted strong hybridization between EuS and Bi$_2$Se$_3$ surface states, in contrast to Ref.~\onlinecite{eremeev2015interface} which found almost no such hybridization and suggested that non-topological surface states due to band bending are at play. Therefore currently there is no comprehensive theoretical picture for the physics of EuS-Bi$_2$Se$_3$ interfaces. 

In this paper, we present a first-principles DFT based investigation of the structural, energetic, electronic, and magnetic properties of such an interface. We consider a periodic (bulk)  EuS(111)/Bi$_2$Se$_3$ heterostructure. Using a non-local exchange-correlation functional, we develop a simple method to model a magneto-composite material from two crystals, EuS(111) and 7 quintuple layers (QLs) Bi$_2$Se$_3$. We do not find any magnetocrystalline anisotropy, in stark contrast to the results obtained in previous theoretical studies which considered a TI interfaced with a thin film. Additionally, we show that after growing EuS on top of TI, the intrinsic electronic structure of 7 QLs Bi$_2$Se$_3$ is preserved and the Dirac cone is downshifted of $\sim$0.4 eV from the Fermi level. This is due to a partial charge transfer between Eu and Se atoms (0.51 electron), which creates a dipole moment at the interface.  The analysis of the electronic band structure reveals that the topological surface state remains gapless, and we do not find an induced magnetic moment on the TI side.  We therefore find no substantial magnetic proximity effect on the TI surface states. We discuss the implications of our findings in the search for a suitable material for QAH effect and topological magnetoelectric effects. 

\begin{figure*}[ht!]
\centering
\includegraphics[width=\textwidth]{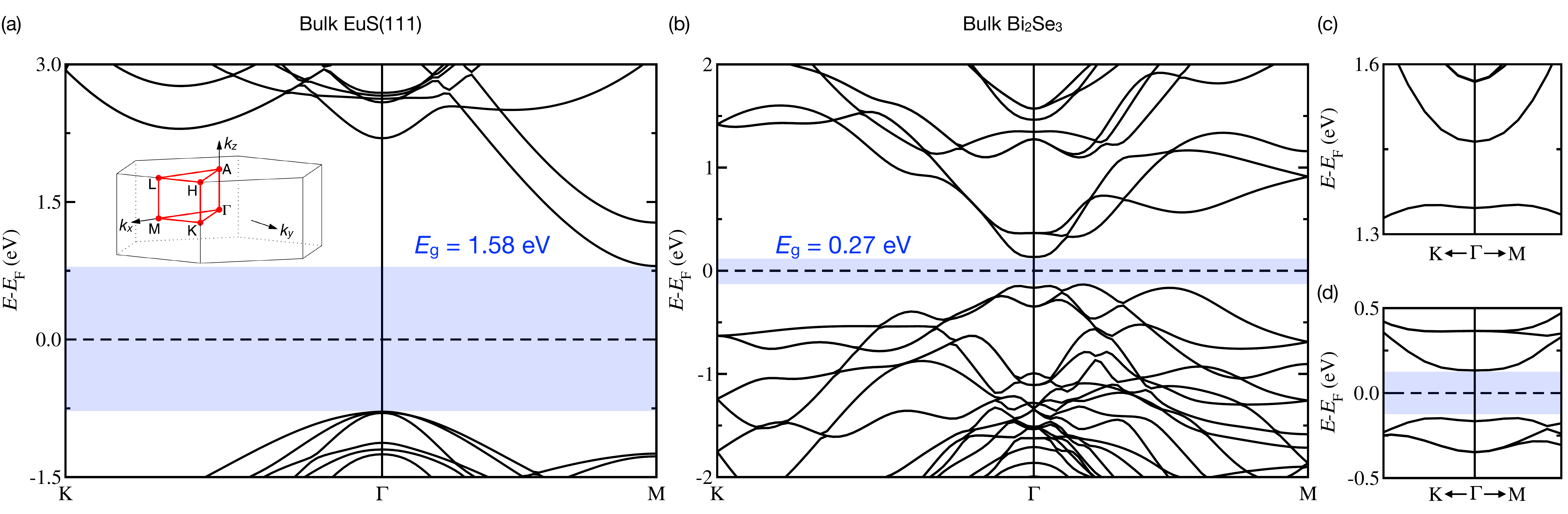}
\caption{Calculated electronic band structure of bulk (a) {EuS}(111) and (b) {Bi$_2$Se$_3$} crystals using PBE+U (with $U_f=10$ eV and $J_f=1$ eV) and DFT-D2 functionals,~\cite{grimme2006semiempirical} respectively. The left inset shows the first Brillouin zone of a hexagonal lattice.~\cite{setyawan2010high} The Fermi level is at zero energy. An indirect electronic band gap, $E_\text{g}=1.58$ eV, is obtained for {EuS}(111), while for {Bi$_2$Se$_3$}, $E_\text{g}=0.27$ eV. Zoom-in electronic band structure of {Bi$_2$Se$_3$} crystal, (c) from 1.3 to 1.6 eV and (d) from -0.5 to 0.5 eV in the vicinity of the $\Gamma$-point.}
\label{Fig-S2}
\end{figure*}

\section{Magnetic proximity effect in topological heterostructures}

EuS material has a rock-salt (NaCl)-type crystal structure with a ferromagnetic (FM) ordered state at low temperature ($T_\text{Curie} = 17$ K), and its electronic structure exhibits a band gap of 1.65~eV.~\cite{jayaraman1974pressure, mauger1986magnetic, ghosh2004electronic} EuS films grow in a (111) orientation with a hexagonal structure that matches well the lattice constant of the (111) surface of Bi$_2$Se$_3$, with the experimental lattice mismatch of 1.86$\%$.

The deposition of EuS(111) film on Bi$_2$Se$_3$ has been extensively studied experimentally as well as theoretically, in an effort to tune the electronic structure and magneto-transport of the TI.~\cite{wei2013exchange,lee2014magnetic,eremeev2015interface,katmis2016high,lee2016direct,kim2017understanding,eremeev2018new,osterhoudt2018charge,krieger2019topology,meyerheim2020structure} Wei \textit{et al.} were the first to measure magneto-transport, observe the QAH effect, and claim the emergence of an FM state in TI.~\cite{wei2013exchange} They concluded that, for an EuS(111) film thickness ranging from 1 to 10~nm, the heterostructure has an in-plane magnetic moment, and even argued for an increased total moment of Eu near the interface compared to the bulk.  More recently, three experimental studies seemed to confirm the presence of both in-plane and out-of-plane magnetic moments close to the interface, presumably due to a strong spin-orbit coupling (SOC) interaction on the TI side.~\cite{katmis2016high,lee2016direct,krieger2019topology} It was even suggested that the FM order in heterostructures persists up to room temperature.~\cite{katmis2016high} However, this effect was comparable for heterostructures of EuS with Bi$_2$Se$_3$ and titanium,~\cite{krieger2019topology} questioning the exact connection to the topological properties. In parallel, Osterhaudt \textit{et al.} performed a series of Raman measurements on EuS(111)/Bi$_2$Se$_3$ heterostructures,\cite{osterhoudt2018charge} and found evidence for a charge transfer between EuS and Bi$_2$Se$_3$ crystals, involving a shift of the Fermi level. In contrast to earlier experiments, two recent studies, one directly measuring X-ray magnetic circular dichroism (XMCD),~\cite{figueroa2020absence} and another, using the high-resolution surface structure X-ray measurements as an input into \textit{ab initio} calculations,~\cite{meyerheim2020structure} found no evidence of induced magnetization in the TI. The latter calculation, somewhat surprisingly, found a large induced moment of sulfur near the interface. 

On the theoretical side, Lee \textit{et al.} used DFT, with a PBE+U scheme and with magnetic moments fixed normal to the interface, to study the electronic structures of 3 and 5 QLs thick Bi$_2$Se$_3$ on top of a EuS(111) film.~\cite{lee2014magnetic} Despite using an unconventional geometry optimization method, relaxing only four atomic layers instead of five in the first QL Bi$_2$Se$_3$, these authors estimate a partial charge transfer from EuS to the TI slab of $\sim$0.5 electron. This implies a downshift of the Dirac point from the Fermi level with a value of 0.25 eV. A similar mechanism has already been observed in doped-graphene.~\cite{tristant2015theoretical,tristant2015theoretical1} Moreover, the authors of Ref.~\onlinecite{lee2014magnetic} noticed that the induced magnetic moment in the TI is very small, as can be expected from the localized $f$-electrons at the heart of Eu magnetic moments, and the penetration of the TI surface states in the EuS is minimal. In addition to the downshift, the Dirac cone next to the interface became gapped, while, for thick enough films, the Dirac state at the opposite film surface remained gapless.

In contrast, Eremeev \textit{et al.}~\cite{eremeev2015interface,eremeev2018new} argued that, after performing a very similar calculation, there is a negligible  magnetic contribution to the gap in the surface states. They also observed a shift of the topological state away from the first QL, and attributed the observed magnetic proximity effect to interdiffusion of Eu in the first QL of the TI. 

To improve the DFT-based description, Kim \textit{et al.} considered~\cite{kim2017understanding} the van der Waals (vdW) correction using DFT-D3 exchange-correlation functional.~\cite{grimme2010consistent} They found an increased magnetic anisotropy in the films relative to the bulk and strong dependence of the anisotropy on the lattice parameters. For the optimized lattice, they showed that the most favorable state is obtained when the magnetization of the EuS film is out-of-plane ($\delta E=0.65$ meV/Eu atom). The magnetic proximity effect in their interpretation was due to significant penetration of the TI surface states into the EuS layers.  

The short literature overview above indicates the conflicting experimental and theoretical results regarding the very existence of the magnetic proximity effect in the EuS/TI heterostructures. Controversy surrounds not only the origin of the energy shift and gapping of the surface states, but also the magnetic anisotropy and the orientation of Eu moments near the interface, as well as mutual interpenetration of the TI surface states and those of the magnetic insulator, and the magnitude of the induced moment. 

To address this problem comprehensively and consistently, we start here a {\em bulk} interface between EuS and Bi$_2$Se$_3$, before moving to the thin film geometry. We report a detailed analysis of the electronic and magnetic properties of such interfaces, as detailed below.

\section{Computational details and methods}

We performed first-principles calculations based on DFT. The structural, electronic, and magnetic properties of all structures were obtained using the Vienna \textit{Ab Initio} Simulation Package (VASP).~\cite{PhysRevB.47.558,PhysRevB.49.14251,PhysRevB.54.11169,KRESSE199615} We used projector augmented wave (PAW) pseudopotentials,~\cite{PhysRevB.50.17953} for Eu $(5s,5p,4f,6s)$, S $(3s,3p)$, Bi $(6s,6p)$, and Se $(4s,4p)$. To account for strong correlation in Eu-$f$ states, we employed GGA+U method of Dudarev \textit{et al.},~\cite{PhysRevB.57.1505} with $J_f = 1$ eV and $U_f = 10$ eV. The particular choice of these values is explained in the electronic supporting information (ESI).~\cite{ESI} A plane-wave basis energy cutoff of 520 eV and a Gaussian smearing of 0.01 eV were found to yield converged total energy and forces. We set the Kerker mixing parameter to 0.01. We have used PBE functional~\cite{kresse1999ultrasoft} including vdW interactions in all our calculations, as they were shown to be important for determining the correct electronic structure of topological and layered materials.~\cite{shirali2019importance,tristant2018finite,kundu2020reversible,sheremetyeva2019first} We investigated vdW-DF, vdW-DF2, optPBE-vdW, optB88-vdW, optB86b-vdW, and SCAN schemes, as well as the semi-empirical DFT-D2 and DFT-D3 methods of Grimme.~\cite{grimme2006semiempirical,klimevs2009chemical,PhysRevB.83.195131,bucko2010improved,grimme2010consistent,sun2015strongly} We found that DFT-D2 provided the most reliable and consistent structural, electronic, and magnetic properties~\cite{ESI} and  used it throughout this study. Relativistic SOC was  included in all calculations. For slab geometries, required to determine the properties of the surfaces, we added a vacuum layer of 100 \AA\ to avoid interaction between periodic images along the $c$-direction. 

We followed a full geometry relaxation (atoms and the cell are relaxed) to a force cutoff of 10$^{-3}$ eV.\AA$^{-1}$, and performed self-consistent calculations until the total energy converges to 10$^{-6}$ eV. For $k$-point sampling we employed a $\Gamma$-centered $(11\times 11\times 11)$ grid for all bulk calculations, and a $\Gamma$-centered $(11\times 11\times 1)$ grid  for slab calculations and the interface systems. To quantify the number of electrons and the local magnetic moment per atom, we carried out a Bader charge analysis.~\cite{henkelman2006fast,sanville2007improved,tang2009grid} 

For the sake of completeness, we have also computed the topological states of the primitive rhombohedral 5-atom unit cell of bulk Bi$_2$Se$_3$, using Quantum Espresso code.~\cite{giannozzi2009quantum,giannozzi2017advanced} We used full relativistic PAW pseudopotentials generated by Kresse-Joubert, \cite{kresse1999ultrasoft} for Bi $(5d,6s,6p)$, and Se $(4s,4p)$. A plane-wave basis energy cutoff of 520 eV (38.2 Ry) and a Gaussian smearing of 0.01 eV ($7\cdot 10^{-4}$ Ry) were set up. We used PBE functional including DFT-D2 scheme and SOC was included. We employed a $\Gamma$-centered $(11\times 11\times 11)$ grid. We also calculated the $\mathbb{Z}_2$ indices and the surface state spectrum were performed using WannierTools code.~\cite{WU2017} For this purpose, maximally-localized Wannier functions based on Bi-$pd$ and Se-$pd$ orbitals were generated \textit{via} the Wannier90 code.~\cite{mostofi2014updated} 

\section{\textit{Ab initio} calculations of heterostructures}
\subsection{Bulk properties} 

To analyze heterostructures we first need to identify the theoretical approaches that yield accurate results for the constituent bulk materials. We performed an extensive analysis of different exchange-correlation functionals for both materials considered. We summarize our findings here (details are provided in the ESI~\cite{ESI}) and compare our results with experimental~\cite{jayaraman1974pressure,mauger1986magnetic,ghosh2004electronic,nakajima1963crystal,zemann1965crystal} and other calculated values~\cite{luo2012first,shirali2020importance}.

For the EuS(111) structure, we carried out DFT+U calculations to describe the localized $f$-electrons that standard DFT fails to treat accurately. We considered a set of $U_f$ and $J_f$ parameters within the GGA+U scheme as implemented in the VASP code, where $U_f$ represents the screened on-site Coulomb interaction, while $J_f$ is the exchange interaction (see Table S1 in the ESI).~\cite{ESI} Using the PBE functional we obtained the ordered moment of 6.96 $\mu_\text{B}$ per Eu atom, essentially independent of the values of $U_f$ and $J_f$, while the lattice constants vary relatively weakly, by 0.5\%, within the considered range of these parameters. In contrast, the energy gap varies by more than a factor of 2, with $U_f=10$ eV and $J_f=1$ eV, yielding the indirect band gap energy of 1.58 eV (see Fig. \ref{Fig-S2}(a)), close to the experimental value of 1.65 eV.~\cite{jayaraman1974pressure,mauger1986magnetic,ghosh2004electronic} For the same parameter values the lattice constants are overestimated by 0.85\% ($a=4.256$ \AA$\,$ and $c=10.424$ \AA). Comparison of several different exchange-correlation functionals (vdW-DF, vdW-DF2, optPBE-vdW, optB88-vdW, optB86b-vdW, DFT-D2, DFT-D3, and SCAN) confirms that the best agreement with experimental data is obtained for semi-local PBE+U method (see Table S2 in the ESI).~\cite{ESI} We also find no magnetic anisotropy in the bulk at optimized structural values.

For the Bi$_2$Se$_3$ material, we also compared the results of a large number of exchange correlation functionals, paying special attention to those that account for the vdW interaction between the QLs (see Table S2 in the ESI).~\cite{ESI} Our findings agree with those of Shirali \textit{et al.}.~\cite{shirali2020importance}, who suggested that the inclusion of empirical vdW forces in DFT-D2 method provides the best agreement with experiment. For this method, we find lattice constants $a=4.135$ \AA$\,$ and $c=28.659$ \AA$\,$ which  differ from the experimental values by $-0.2$\% and $0.1$\% respectively. Furthermore, the obtained indirect band gap energy of 0.27 eV (see Fig. \ref{Fig-S2}(b))is in good agreement with the experimental value of 0.30~eV.~\cite{nakajima1963crystal,zemann1965crystal} We also find, somewhat surprisingly, that the vdW-DF, vdW-DF2, optPBE-vdW, optB88-vdW, and optB86b-vdW significantly overestimate the $a$ and $c$ parameters, by up to 7.3\% (4.447 \AA) and 36.5\% (39.076 \AA), respectively. These large deviations are only obtained when the relativistic SOC effects are included.~\cite{ESI} For example, in the case of optB88-vdW {\em without relativistic effects}, the same values are $a=4.192$ \AA$\,$ (0.6\%) and $c=29.213$ \AA$\,$ (-1.0\%). We cannot judge whether this is an intrinsic problem or occurs due to subtleties of the implementation of these relativistic calculations in the VASP code. 

\begin{figure}[t!]
\centering
\includegraphics[scale=0.23]{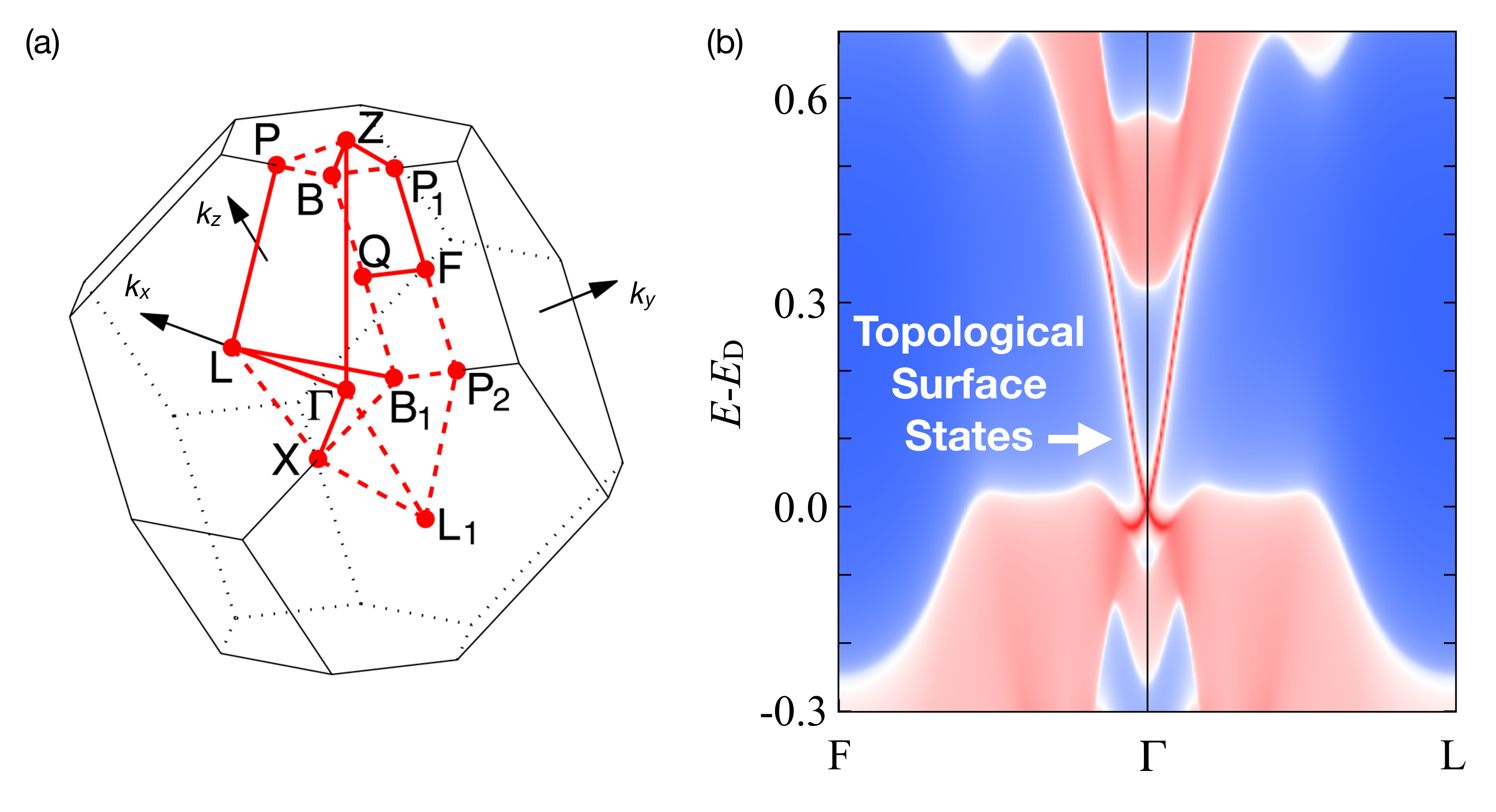}
\caption{(a) The first Brillouin zone of a rhombohedral lattice.~\cite{setyawan2010high} (b) Calculated surface state spectrum from bulk Bi$_2$Se$_3$ rhombohedral unit cell, using DFT-D2 functional and spin-orbit coupling. \cite{grimme2006semiempirical} The energy region of bulk-projected and topological surface-projected states are indicated by light and dark red colors, respectively. The Dirac cone is at zero energy.}
\label{Fig-6}
\end{figure}

To further validate the use of DFT-D2, we calculated the topological features of the Bi$_2$Se$_3$ material. Since this insulating material has inversion symmetry, the four $\mathbb{Z}_2$ topological indices, \textit{i.e.} $\nu_0;(\nu_1\nu_2\nu_3)$, can be computed as the product of half the parities at each of the high-symmetry points in the first Brillouin zone (BZ) of a rhombohedral lattice (see Fig.~\ref{Fig-6}(a)).~\cite{moore2007topological,fu2007topological,roy2009topological} With this notation, $\nu_0=1$ indicates a strong TI, for $\nu_0=0$ and any non-zero $\nu_i$ with $i=\{1,2,3\}$, denotes a weak TI, while a trivial insulator is defined by $\nu_i=0$ $\forall i$. We calculated these indices from the evolution lines of Wannier charge centers for bulk Bi$_2$Se$_3$ (see Fig. S3 in the ESI).~\cite{ESI} At the $\Gamma$-point, the evolution lines indicate that the $\mathbb{Z}_2$ topological indices are $1;(000)$. Then, we use the surface Green's function formalism to calculate the surface state spectrum from bulk Bi$_2$Se$_3$. Figure ~\ref{Fig-6}(b) shows that the surface state is localized in the bulk band gap energy in the vicinity of the Fermi level (dark red lines). These results are consistent with data found in the literature,~\cite{yu2011equivalent} and the calculated electronic band structures of multi-QLs Bi$_2$Se$_3$ slabs reported in Fig.~S2 in the ESI.~\cite{ESI} They confirm that the use of DFT-D2 scheme does not alter the strong topological properties of bulk Bi$_2$Se$_3$.

\subsection{EuS(111)/Bi$_2$Se$_3$ interface}

\begin{figure*}[t!]
\centering
\includegraphics[width=\textwidth]{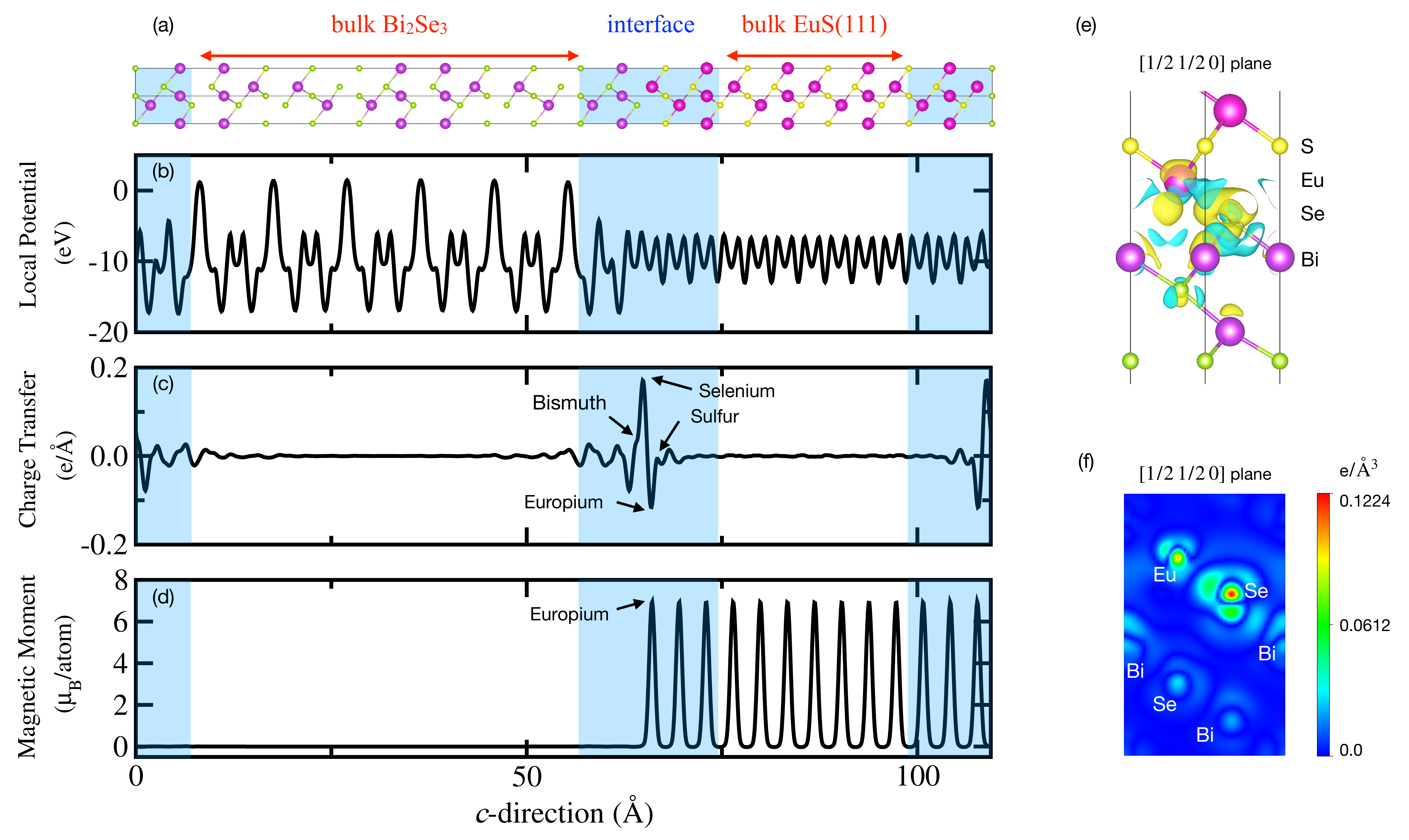}
\caption{(a) Side view of EuS(111)/ 7 quintuple layers Bi$_2$Se$_3$ heterostructure with $a_{\text{EuS(111)}}=4.256$ \AA$\,$ and out-of-plane magnetization. Eu, S, Bi, and Se atoms are represented in pink, yellow, purple, and green, respectively. The blue shaded part represents the interface zone. (b) Calculated local potential, (c) charge transfer per unit cell area, and (d) local magnetic moment as a function of the \textit{c}-direction. (e) Charge density transfer at the interface. Yellow represents positive electronic densities which accept electrons and cyan represents negative densities which donate electrons. (f) Two-dimensional charge density transfer at the interface, projected onto the $(1/2\, 1/2\, 0)$ plane.}
\label{Fig-1}
\end{figure*}

We are now in the position to describe the interface between bulk materials. From this point onward each reference to EuS corresponds to the hexagonal structure oriented in the (111) direction unless indicated otherwise. We consider a periodic heterostructure of 7 QLs Bi$_2$Se$_3$ and 25 atomic layers of EuS, depicted in Fig.~\ref{Fig-1}(a). This is in contrast to much of earlier work that considered interface slab models with a vacuum region. As with all periodic structures, there are two topological interfaces, and our choice of material thickness ensures that they do not significantly interact electrostatically, and that the topological interface states are not hybridized, see below. 

Our results above showed that the PBE+U ($U_f=10$ eV and $J_f=1$ eV) and the non-local DFT-D2 functionals are the best options to describe the properties of EuS and Bi$_2$Se$_3$ crystals, respectively. Therefore at the interface one would expect the need for a tailored DFT-D2 functional to change the potential smoothly, by gradually varying the values of the vdW dispersion coefficients of each atom in the first QL Bi$_2$Se$_3$. We  compared this result to that obtained using the standard non-local functional, and found no sudden change in potential at the interface for both calculations. Therefore, we did not tailor the functional for the results shown in this study. The resulting potentials are plotted in Fig.~\ref{Fig-1}(b), and discussed below.

\begin{figure}[t!]
\centering
\includegraphics[scale=0.27]{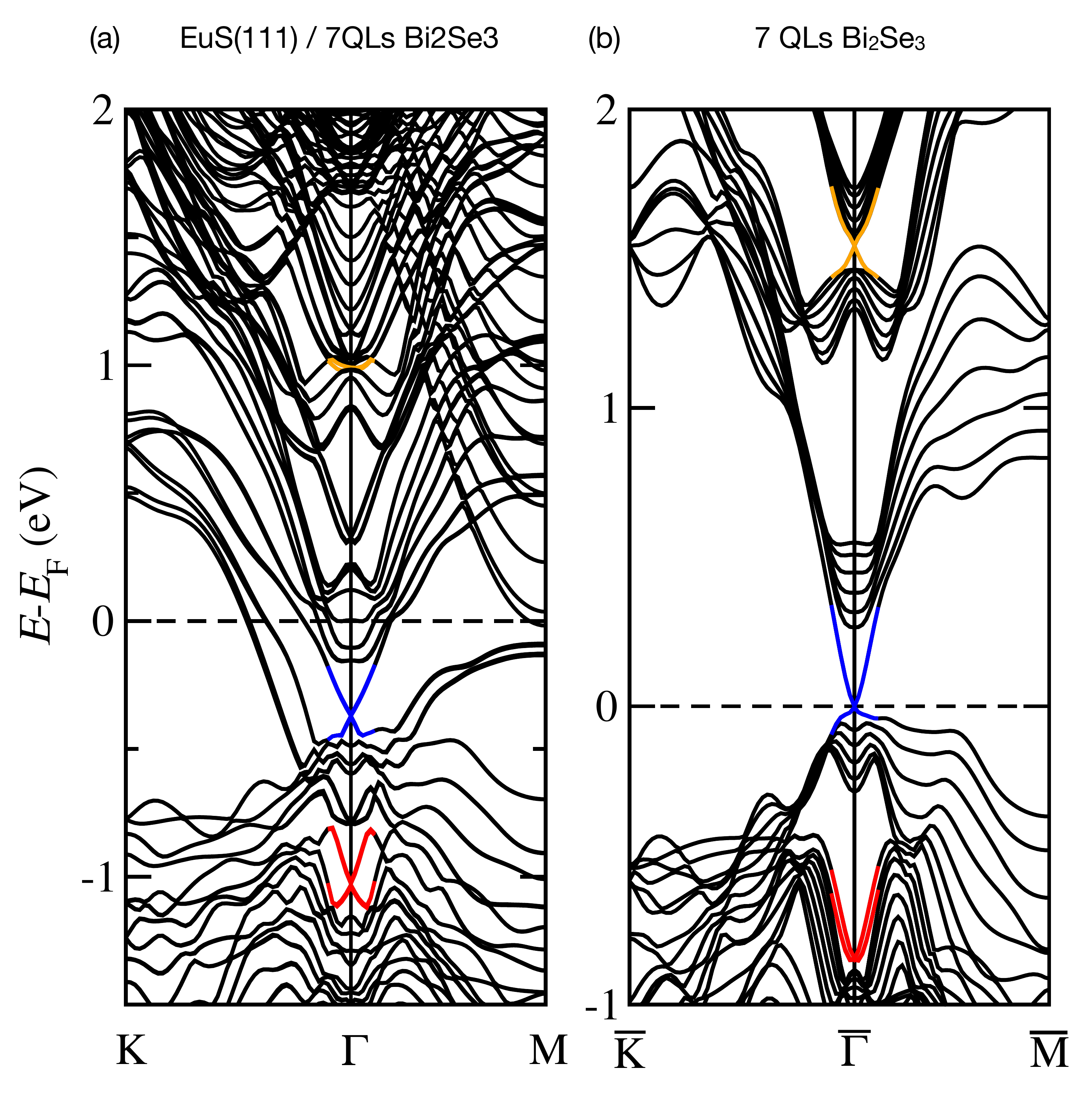}
\caption{Calculated electronic band structure of (a) EuS(111)/ 7 quintuple layers (QLs) Bi$_2$Se$_3$ and (b) Bi$_2$Se$_3$ slab with 7 QLs. The Fermi level is at zero energy. The non-topological Rashba interface state below the chemical potential, the topological interface state, and the quasi-linear high energy state are in red, blue, and orange, respectively.}
\label{Fig-2}
\end{figure}

To analyse the magnetic properties of an interface, we considered four configurations. We set the in-plane lattice constants equal to the bulk EuS ones, $a_{\text{EuS(111)}}=4.256$ \AA$\,$ (lattice mismatch $\delta_{\text{EuS(111)}}=2.93$\%), with (i) out-of-plane and (ii) in-plane magnetization, and that optimizing bulk Bi$_2$Se$_3$,  $a_{\text{Bi$_2$Se$_3$}}=4.135$ \AA$\,$ ($\delta_{\text{Bi$_2$Se$_3$}}=-2.84$\%), again with (iii) out-of-plane and (iv) in-plane magnetization. 

\begin{figure*}[t!]
\centering
\includegraphics[scale=0.30]{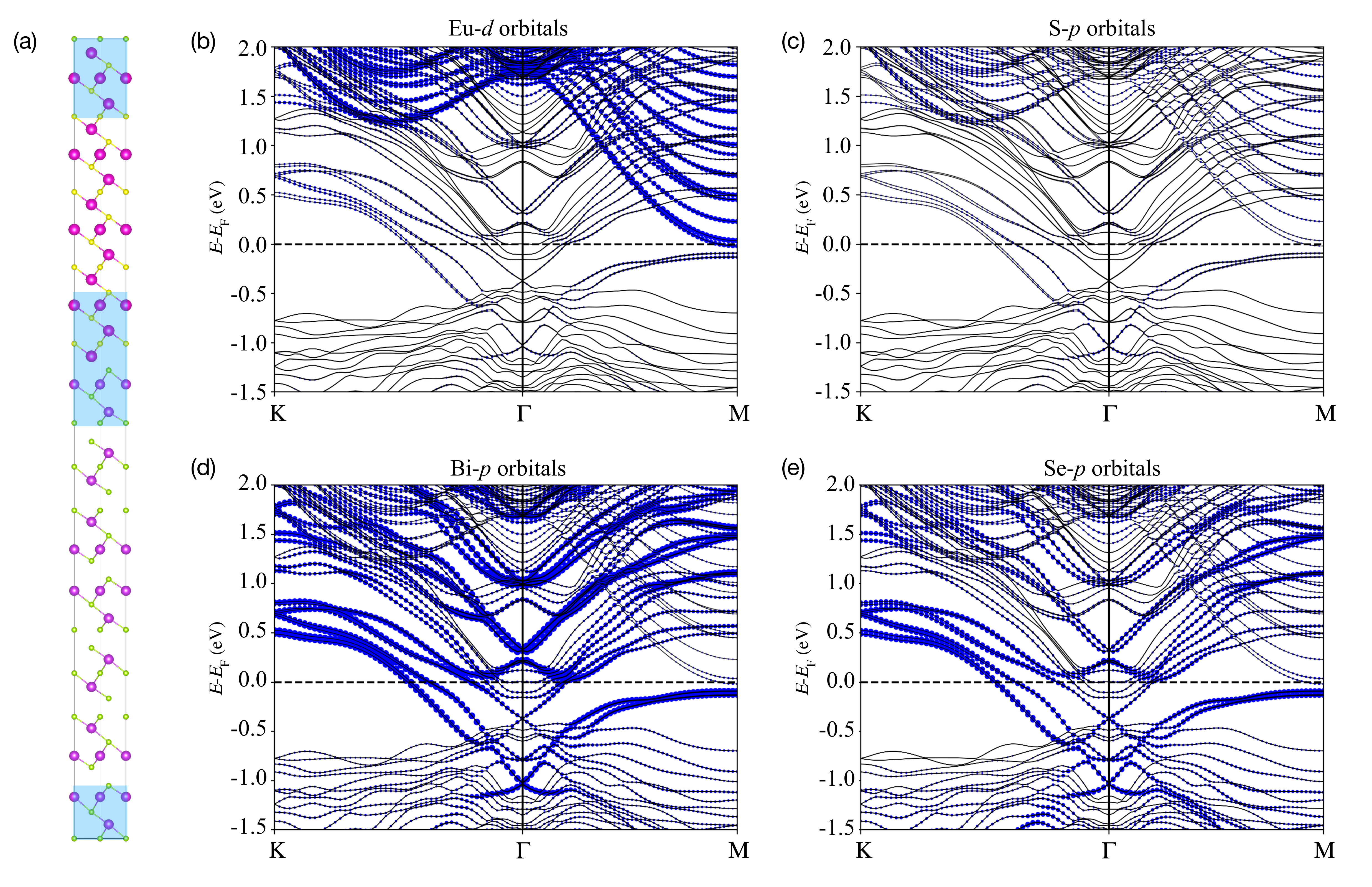}
\caption{Atom-projected band structure at the interface of (a) EuS(111)/ 7 quintuple layers Bi$_2$Se$_3$ onto (b) Eu-$d$, (c) S-$p$, (d) Bi-$p$, and (e) Se-$p$ orbitals. The size of the filled circles are proportional to the weight projected onto the orbitals. The Fermi level is at zero energy. The blue shaded part represents the interface zone.}
\label{Fig-4}
\end{figure*}

\begin{figure*}[t!]
\centering
\includegraphics[scale=0.28]{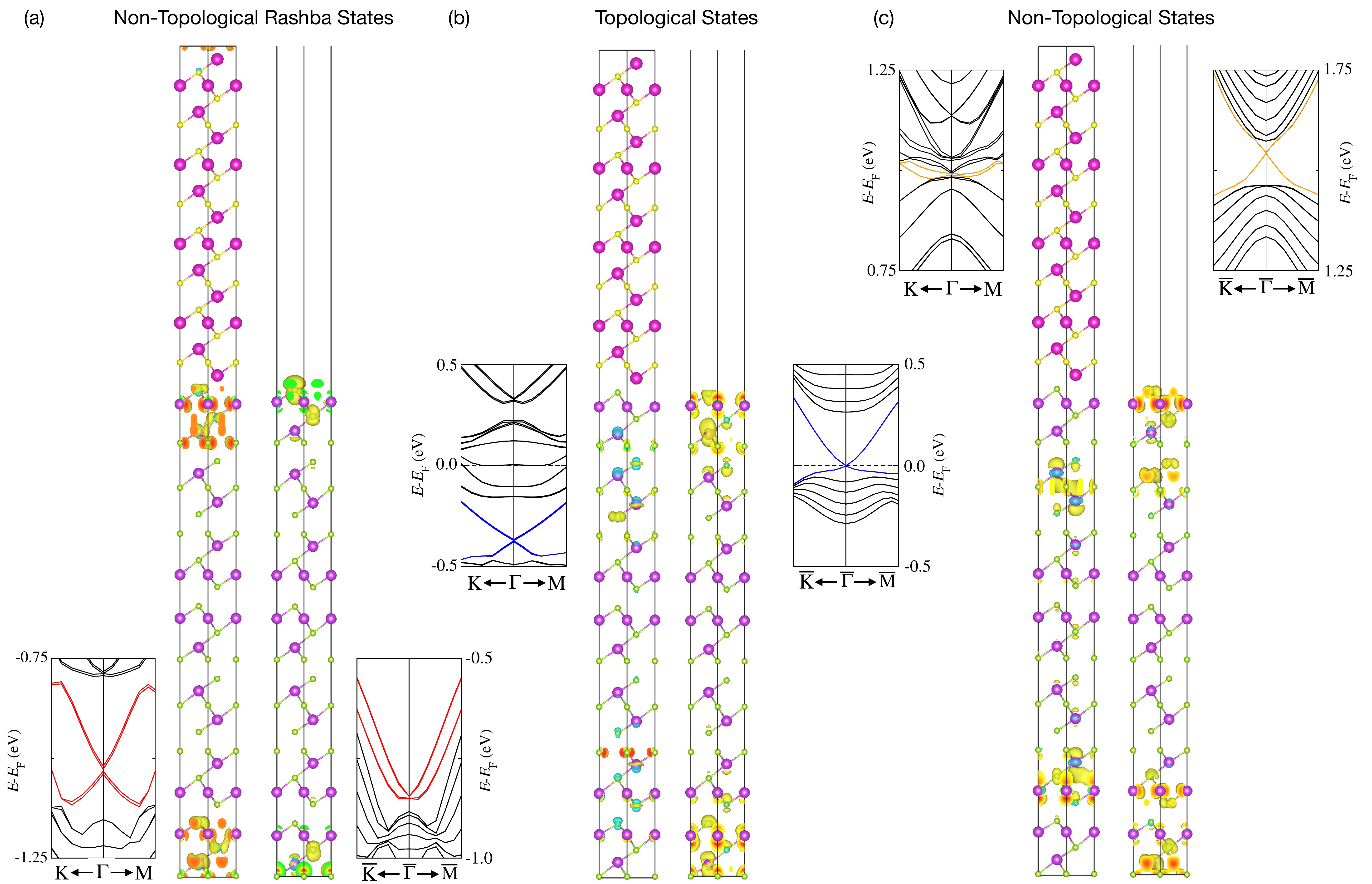}
\caption{Electronic band structure and band-decomposed charge density distributions at the $\Gamma$-point of EuS/Bi$_2$Se$_3$ and 7 quintuple layers Bi$_2$Se$_3$ for (a) non-topological Rashba interface states below the chemical potential (red lines), (b) topological interface states (blue lines), and (c) quasi-linear high energy states (orange lines). The Fermi level is at zero energy.}
\label{Fig-3}
\end{figure*}

For a set of fixed in-plane lattice constants we still need to determine the optimal distance between the two materials at the interface. In a computational simulation, the optimized interfacial region thickness is calculated by tuning the length of the structure in the $c$-direction, normal to the interface. For several unit cell volumes, we determine the minimum ground state energy and the interlayer distance between both materials. We allow the first QL in the TI slab and six atomic layers in the EuS slab closest to the interface (blue shaded in Fig.~\ref{Fig-1}) to relax while the rest of the atoms are fixed. Due to the periodicity, we carefully checked that the Eu-Se bond lengths, at both interfaces, are identical. The ground state energy as a function of the length $c$ is then fit with a third degree polynomial, as shown in Fig.~S4 in the ESI.~\cite{ESI} Finally, from the fitted function, we extracted the out-of-plane lattice parameter that minimizes the ground state energy of each system (red square) and calculated their optimized parameters, which are summarized in Table~\ref{Tab-1}. For $a_{\text{EuS(111)}}$, the Eu-Se bond length turns out to be similar to the experimental value for bulk  EuSe rock-salt structure ($a=3.095$ \AA),~\cite{jayaraman1974pressure} while it tends to be slightly underestimated by $\sim$0.02 \AA$\,$ for $a_{\text{Bi$_2$Se$_3$}}$. 

Next, the magnetic anisotropy energy can be directly obtained by calculating the difference in ground state energy between in-plane and out-of-plane magnetizations. We did not find any significant magnetic anisotropy ($\delta E = 0.1$ meV, \textcolor{black}{same order of magnitude than Ref.~\cite{kim2017understanding}}) for either in-plane lattice constant ($a_{\text{EuS(111)}}$ and $a_{\text{Bi$_2$Se$_3$}}$), indicating that both magnetic structures can coexist at zero temperature. Therefore the magnetic structure realized in a specific experiment may depend on the details of the synthesis, thereby potentially explaining the different results obtained in experiments.~\cite{katmis2016high,lee2016direct,krieger2019topology}

\begin{table}[b!]
\caption{Eu-Se bond length (\AA), ground state energy per atom (eV), and local magnetic moment ($\mu_\text{B}$/Eu atom), obtained for two in-plane lattice constants ($a_{\text{EuS(111)}}=4.256$ \AA$\,$ and $a_{\text{Bi$_2$Se$_3$}}=4.135$ \AA) and two spin orientations (in-plane and out-of-plane). Spin-orbit coupling is included.}
\centering
\begin{tabular}{l l l l l l}
\hline
\hline
Configuration & ($a$,$b$) & Spin & Eu-Se & Energy & $m_{\text{spin}}$ \\
\hline
(i)   & EuS(111)     & out-of-plane & 3.091 & -6.5976  &  6.99 \\
(ii)  & EuS(111)     & in-plane     & 3.090 & -6.5976  &  6.99 \\
(iii) & Bi$_2$Se$_3$ & out-of-plane & 3.072 & -6.5959  &  6.99 \\
(iv)  & Bi$_2$Se$_3$ & in-plane     & 3.071 & -6.5960  &  6.99 \\
\hline
\hline
\end{tabular}
\label{Tab-1}
\end{table}

Further insight is gained by analyzing the local potential as a function of atomic position along the $c$-direction, shown for configuration (i) in Fig.~\ref{Fig-1}(b), where the blue shaded area indicates the interface region that has been relaxed in the calculation. Approximately 10~\AA$\ $ away from the interface, the potentials of EuS and Bi$_2$Se$_3$ already converge to the bulk value: absence of drift in the height of the peaks throughout the white region, where the atomic positions are held fixed, indicates no sensitivity to the interface region. The deviations from this behavior are constrained to the immediate vicinity of the interface, and we attribute it to charge redistribution across the region where the two material systems interact. 
We show the plane-averaged charge density difference along the \textit{c}-direction in Fig.~\ref{Fig-1}(c). The presence of the interface induces a partial charge transfer of $\sim$0.51~electron from Eu to Se atoms. This charge transfer leads to a dipole moment, which can be seen by projecting the result onto the $(1/2\, 1/2\, 0)$ plane (see Figs.~\ref{Fig-1}(e) and \ref{Fig-1}(f)). Finally, the local magnetic moments of each atom are also presented in Fig.~\ref{Fig-1}(d). As may be expected, the spins of all Eu atoms remain close to 7 $\mu_\text{B}$, since the partially filled $4f$ orbitals, where the moment originates, are located $\sim$2 eV below the Fermi level. As a consequence, these electrons are insensitive to the weak perturbations of the potential. Moreover, due to the lack of hybridization between Eu and Bi states, no induced magnetization within the first QL Bi$_2$Se$_3$ is found. This finding is in contrast to the results obtained in previous theoretical studies,~\cite{lee2014magnetic,eremeev2015interface} but it is in agreement with recent experimental measurements.~\cite{figueroa2020absence} 

To confirm our findings, we investigated a heterostructure where 6 layers of EuS were stacked on top of a 5 QLs Bi$_2$Se$_3$ slab where the S atoms at the EuS surface are passivated with hydrogen atoms. We also included a vacuum region of 50 \AA. The results are described in detail in Fig.~S6 in the ESI.~\cite{ESI} We found a partial charge transfer of $\sim$0.52~electron from Eu to Se atoms, and the magnetic moments of all Eu atoms remain close to 7~$\mu_\text{B}$. These results are almost identical to those obtained for the periodic heterostructure, and this comparison strongly confirms the conclusion that the interfacial properties of EuS/Bi$_2$Se$_3$ are extremely localized, and are almost insensitive to EuS film thickness.

\subsection{Electronic structure at the interface}

So far, we have established that the presence of an interface does not induce magnetic anisotropy and that the essential physics is governed by the charge transfer. We will now study the consequences of the electrostatic effects on the electronic structure of EuS/Bi$_2$Se$_3$ heterostructure (Fig.~\ref{Fig-2}(a)). For comparison, we also show the band structure of 7 QLs Bi$_2$Se$_3$ slab in Fig.~\ref{Fig-2}(b), as it allows us to provide a connection between the surface and the interface states. This comparison is based on the atom- and orbital-projected band structure at the interface of EuS/Bi$_2$Se$_3$, displayed in Fig.~\ref{Fig-4}. As seen in that Figure, Eu-$d$- and S-$p$-derived bands (around $~$1.5 eV at the $\Gamma$-point) do not hybridize with Bi-$p$ and Se-$p$ states in the vicinity of the Fermi level. As a result, the intrinsic electronic structure of Bi$_2$Se$_3$ surface is preserved in the heterostructure. However, the energy is shifted due to the charge redistribution. It is noteworthy that this is a consequence of the insulating nature of EuS, and the same features are not observed when TIs are interfaced with doped substrates or metallic ferromagnets.~\cite{checkelsky2012dirac,baker2015spin,wang2017room}

Fig.~\ref{Fig-2}(a) shows that the EuS/Bi$_2$Se$_3$ interface exhibits a metallic behavior. This metallicity is due to the Bi- and Se- electrons. This can be seen directly by inspecting Figs.~\ref{Fig-4}(d) and \ref{Fig-4}(e) where the corresponding bands are found to cross the chemical potential nearly halfway between $K$ and $\Gamma$ points in the BZ. Consequently, we will now focus on the vicinity of the $\Gamma$-point to identify the distinguishing features of the heterostructure inherited from Bi$_2$Se$_3$. 

Our study will focus on three states highlighted using different colors in Fig.~\ref{Fig-2}: the topological state (blue bands), linear crossing above the chemical potential (orange bands), and the Rashba-split pair of states below the Fermi energy (red bands).  Near the $\Gamma$-point, the topological interface state in the  EuS/Bi$_2$Se$_3$ heterostructure is downshifted by $\sim$0.4 eV compared to its position near the chemical potential in the  Bi$_2$Se$_3$ slab, due to the interfacial charge transfer. Since there is no induced magnetization within the first QL of Bi$_2$Se$_3$, see Fig.~\ref{Fig-1}(d), this degenerate Dirac cone remains gapless and essentially intact. The induced dipole moment does not affect the cone's properties, except by renormalizing the Dirac velocity from the computed value of $5.07\times 10^5$m/s for the slab of Bi$_2$Se$_3$ (in agreement with Ref.~\cite{shirali2019importance} which employed the same methods, and with the experimental results) to a smaller $v_\text{D}\approx 2.73 \times 10^5$m/s at the interface. 

We note that the surface state is sensitive to the number of TI layers. For instance, Lee \textit{et al.} reported that gaps of 80 and 10 meV at the $\Gamma$-point arise in this topological state when EuS is interfaced with 3 and 5 QLs Bi$_2$Se$_3$, respectively.~\cite{lee2014magnetic} In the present case, 7 QLs represent a threshold for the closing of this gap which we believe to be in part driven by the hybridization at opposite surfaces. As discussed below, we find that the topological states move deeper into the Bi$_2$Se$_3$ layer as a consequence of the charge redistribution, and therefore the hybridization gap is more pronounced for the heterostructures than for Bi$_2$Se$_3$ surfaces.

We also track two other notable features of the electronic band structure of Bi$_2$Se$_3$ slab. First, a linear crossing point appears at the same high-symmetry point in the first BZ, but $\sim$1.5 eV above the Fermi energy (orange bands), as already noted in Refs.~\onlinecite{sobota2013direct,zhu2015effect}. These states are absent in the bulk Bi$_2$Se$_3$, as seen in Fig. \ref{Fig-S2}(c). In the heterostructure, these surface states are downshifted $\sim$0.5 eV, and lose their linear dispersion as they become sandwiched between the conduction bands (Fig.~\ref{Fig-2}). 
Second, an interface state that is Rashba-split due to the combined effect of SOC interactions and asymmetry of the crystal potential (red lines in Fig.~\ref{Fig-2}) is visible at the $\bar{\Gamma}$-point, $\sim$0.80 eV below the Fermi level in Bi$_2$Se$_3$ slab.~\cite{bihlmayer2015focus} This state is also downshifted by $\sim$0.30 eV in the EuS-Bi$_2$Se$_3$ heterostructure, and the Rashba splitting is increased, resulting in quasi-linear crossing. Therefore our analysis suggests a consistent downshift in energy of the main interface states, while preserving their essential properties. 

Intriguingly, while all three states that we focused on reside in the surface QL of a Bi$_2$Se$_3$ slab, they respond differently to the formation of the interface. Fig.~\ref{Fig-3} shows the electronic density of these states for such a slab compared to the interface. The Rashba state remains mostly localized in the topmost QL, which is consistent with the greater splitting observed in our results. On the other hand, due to charge screening, both the high-energy state and the topological interface state are pushed towards the second QL (and possibly even further away from the interface for the topological state, see Fig.~\ref{Fig-3}(b)). This helps to understand the lack of magnetic proximity coupling to EuS, as the overlap between the wave function of the topological state and that of the Eu $f$-shell electrons is reduced. \textcolor{black}{Reduction of the overlap between the EuS and topological states due to the “submergence” of the latter into the second QL was invoked  to explain the weakness of magnetic proximity effect.~\cite{eremeev2013magnetic,men2013magnetic}} It is also consistent with the picture where the hybridization between those states in enhanced for thin TI films, resulting in observable gaps. However, we emphasize that according to our results, these hybridization gaps are not caused by the magnetic proximity coupling to the EuS spins.

Our results above have been presented for the configuration (i) in Table~\ref{Tab-1}. For the sake of comparison, we also calculated the electronic band structure of EuS/Bi$_2$Se$_3$ by considering configurations (ii), (iii), and (iv). These results are shown in Fig~S5 in the ESI.~\cite{ESI}  
All four cases predict the three surface features mentioned above at the same position. We therefore conclude that neither the spin orientation nor the choice of one of the materials (EuS or Bi$_2$Se$_3$) as the substrate  affect the electronic properties of EuS/Bi$_2$Se$_3$ heterostructure, and our conclusions remain robust.

\section{Discussion and conclusions}

We studied the magnetic and electronic structure properties of periodic EuS(111)/Bi$_2$Se$_3$ heterostructures \textit{via} first-principles calculations, including both van der Waals corrections and relativistic spin-orbit coupling effects. In contrast to previous theoretical predictions, we found that this system does not exhibit any significant magnetocrystalline anisotropy. The topological surface state is localized in the second QL from the interface, is gapless, and is downshifted $\sim$0.4 eV from the Fermi energy. These properties results from a partial charge transfer of 0.51 electron from Eu to Se atoms, which creates a dipole moment at the interface and, as a result,  downshifts of the Bi$_2$Se$_3$-$p$ bands. In addition, since Eu magnetization emerges from the partially filled $4f$ orbitals, which are located 2 eV below the Fermi level, there is no substantial magnetic proximity effect on the electronic properties of Bi$_2$Se$_3$. These properties are consistent with recent experimental results.~\cite{krieger2019topology, figueroa2015local} 

Our results bring into focus the challenges associated with achieving a significant proximity coupling between magnetic materials and topological insulators. We note that the growth condition may have a significant effect both on the sharpness of the interface, which affects the electronic properties,~\cite{eremeev2018new} and on the magnetic structure, due to absence of magnetic anisotropy in our results. This may explain the variability of experimental results in the literature. Controlling the growth conditions to realize the most favorable structure for the magnetic proximity effect on the topological interface state is difficult, but may be a viable pathway for the heterostructures containing localized magnetic moments. 

Another possible path forward is to select, instead of EuS, magnetic compounds where the electronegativity of the atoms across the interface is comparable. This would presumably not lead to such dramatic charge redistribution and would leave the topological states localized in the first structural unit, and possibly leak into the magnetic material sufficiently to couple to the spin degrees of freedom. Yet another alternative may be to rely on magnetism derived from the less localized electrons in $d$-shells. In principle, in ultrathin films of Bi$_2$Se$_3$ there may be sufficient repulsion between the topological states at the opposite sides of the film to enhance the leakage into the magnetic materials. The flip side of this approach is, of course, the gap opening due to the hybridization of these states. All of these possibilities are worth exploring, and our results provide a pathway towards \textit{ab initio} modeling of such interfaces.

\begin{acknowledgments}
D. T. and W. S. acknowledge support from the US Department of Energy under EPSCoR Grant No. DE-SC0012432 with additional support from the Louisiana Board of Regents. I. V. acknowledges support from NSF Grant No. DMR 1410741. Part of this work was performed using supercomputing resources provided by the Center for Computation and Technology (CCT) at Louisiana State University.
\end{acknowledgments}

\nocite{*}

\bibliography{reference}

\providecommand{\noopsort}[1]{}\providecommand{\singleletter}[1]{#1}%

\end{document}